\documentclass[12pt,sort&compress]{article}

\usepackage{amsmath}
\usepackage{mathtools}
\usepackage{amsfonts}
\usepackage{setspace}
\usepackage{subfigure}
\usepackage{authblk}
\usepackage[numbers,sort]{natbib}

\def\L{\left}
\def\R{\right}

\newcommand{\Dc}[2]{\L(\prescript{C}{}{\mathbf{D}}_{\bar{T}(p_0)}^{#2} #1 \R)\L(t\R)} 

\begin{document}

\title{A physically based connection between fractional calculus and fractal geometry}

\author[1]{Salvatore Butera\thanks{sb469@hw.ac.uk}}
\author[2]{Mario Di Paola\thanks{mario.dipaola@unipa.it}}

\affil[1]{SUPA, Institute of Photonics and Quantum Sciences, Heriot-Watt University, Edinburgh, EH14 4AS, United Kingdom.}
\affil[2]{Dipartimento di Ingegneria Civile Ambientale, Aerospaziale e dei Materiali (DICAM), Universit\`{a} degli Studi di Palermo, Viale delle Scienze,Ed.8, 90128 - Palermo, Italy}

\date{\copyright\ 2015. This manuscript version is made available under the CC-BY-NC-ND 4.0 license http://creativecommons.org/licenses/by-nc-nd/4.0/} 

\maketitle


\begin{abstract}
We show a relation between fractional calculus and fractals, based only on physical and geometrical considerations. The link has been found in the physical origins of the power-laws, ruling the evolution of many natural phenomena, whose long memory and hereditary properties are mathematically modelled by differential operators of non integer order. Dealing with the relevant example of a viscous fluid seeping through a fractal shaped porous medium, we show that, once a physical phenomenon or process takes place on an underlying fractal geometry, then a power-law naturally comes up in ruling its evolution, whose order is related to the anomalous dimension of such geometry, as well as to the model used to describe the physics involved. By linearizing the non linear dependence of the response of the system at hand to a proper forcing action then, exploiting the Boltzmann superposition principle, a fractional differential equation is found, describing the dynamics of the system itself. The order of such equation is again related to the anomalous dimension of the underlying geometry.
\end{abstract}



\section{Introduction}
It is widely believed that the origin of the concept of fractional derivative dates back to the 1695, when Leibniz, in its correspondence with de L'Hospital, posed the question about the meaning of a derivative of order one half. That intuition represented the seed for mathematicians of the following three centuries as Liouville, Riemann, Gr\"unwald, Letnikov et al., to develop the theory of the fractional calculus as it is nowadays known. However, despite of its long time origins, only in the recent few decades the fractional calculus gained the attention of the scientific community, having been considered, for most of the time, as a pure theoretical field of research. The rise of the interest on the fractional calculus, has to ascribe to its suitability in modelling a large variety of phenomena, ranging almost all the fields of the naturals sciences. Nowadays, its applications can be found in biology and biomechanics \cite{Xu2009,Arafa2012,Ahmed2012,Deseri2013,Magin2010a,Magin2010b}, in modelling the rheological properties of viscoelastic materials \cite{MainardiSurvey,Koeller1984,Bagley1986}, in electrical circuits \cite{Ala2014}, in anomalous transport and diffusion processes in complex media \cite{Schneider1989, Isichenko1992,Zaslavsky2002, Metzler2001}, and in many other branches of physics and engineering \cite{podlubny1998,kilbas2006,mainardi2010}. The powerful of the fractional operators relies on their non-local power-law structure, that makes them suitable in describing the memory and hereditary properties of many physical phenomena and processes, and in particular in modelling the ``anomalous'' dynamics of complex systems, where scaling power-laws emerge as a consequence of such complexity. It has been shown in fact, that fractional (integro-)differential equations naturally arise once systems characterized by power law long range interactions, or by power law memory, are considered \cite{Tarasov2008,Cottone2009,DiPaola2008}. Despite of their widely recognized usefulness in describing nature, never a clear and universally accepted geometrical interpretations has been pointed out for such operators.

The development of the theory of fractals, formulated by Mandelbrot in the second half of the twentieth century \cite{mandelbrot1983}, who consolidated the work of many other prior mathematicians and scientists in general, opened a new perspective to this aim. Since that, several attempts have been made, in order to elucidate a connection between the fractional operators and this new geometry, whose anomalous non-integer dimension seemed to be the key of the story. The concept of \textit{fractance} has been one among these attempts. By fractance it is usually referred in literature, a mechanical (or electrical) model, in which classical elements such as springs and dashpot in the case of viscoelasticity (or resistors and capacitors in electrical circuits) are disposed as a fractal tree, with an opportune power law scaling postulated for the relevant physical properties, in order to obtain a constitutive law of non-integer order for the whole system \cite{DiPaola2013,DiPaola2013RS, Deseri2013, Ala2014}. Another line of research has been developed by Nigmatullin and Le Mehaute \cite{Nigmatullin2005,lefleches}, who showed that the procedure of averaging of a smooth function over a non dense self-similar set, leads to a Riemann-Liouville integral operator, whose order is related to the anomalous dimension of the set itself. Working in the time domain, they proposed the interpretation of their result as manifestation of a partial (fractal) memory of the system. A different approach, that led to the definition of a new kind of operators, acting on a non dense support and known in literature as local fractional operators, has been pursued by Kolwankar and Gangal \cite{Gangal2009,Kolwankar1999,Kolwankar2013,Kolwankar1998,Kolwankar1997,Kolwankar1996,Kolwankar1998_2}. The idea underlying their work is that, as they are defined on a continuum support, the ordinary fractional operators cannot be defined on a non dense object like a fractal. This approach, however, required the definition of proper local operators, different from the classical Riemann-Liouville (R-L) or Caputo's fractional derivatives and integrals, missing then the aim of giving a connection between fractal geometry and ordinary fractional calculus. Many others attempts have been proposed in the last decades. The property of fractional operators of changing the anomalous dimension of a random (or deterministic) fractal object, to which they are applied, has for example been shown in \cite{tatom1995}. A similar result was also derived by Rocco and West in \cite{Rocco1999}, where fractal functions have been shown to have fractional derivatives and integrals, whose anomalous dimension differs from the original one. Moreover, by following a different approach, the authors recovered the definition of a critical order of derivation and integration, deduced by Kolwankar and Gangal in \cite{Kolwankar1997,Kolwankar1996}. However, in spite of such effort, never a clear relation between fractals and fractional calculus, only based on geometrical considerations, has been pointed out up to now.

It is our belief, that such a link has not to be sought in the re-definition of fractional operators on fractal support, or in the construction of ad-hoc mechanical (or electrical) models, but in the physical origins of the scaling power laws that emerge from the dynamics of a large variety of complex systems, that are at the basis of the long memory and hereditary properties commons to many natural phenomena, and that is the main peculiarity of fractional operators. In this paper, we want to show that such power laws come out in a natural way, when physical phenomena taking place on an underlying fractal structure are considered. It turns out that the order of such power laws is strictly related to the dimension of the geometry on which the phenomena take place, as well as to parameters related to the physics of the problem. The link between fractals and the raising of dimensional-related power laws is not new in literature and has been investigated in the past, with particular emphasis in diffusion phenomena in complex media \cite{Procaccia1985,havlin1987,Metzler1995,Sokolov2012}. It has been shown for example that, modelling the dynamics of diffusion processes in terms of a random walk, the time dependence of the mean square displacement away from the initial position, deviates from the linearity predicted by the Fick's law for diffusion in homogeneous media, and a power-law of non integer order, related to the fractal dimension of the medium, has been found.

The reason of the emergence of such power laws has to be sought in the definition of fractals as self-similar structures, that differs from regular geometric shapes in their fractional dimensional scaling. In other words, a (deterministic) fractal can be thought as an object, each part of which is a scaled copy of the whole, that differ from a regular geometrical shapes in its non integer scaling coefficient. If we rescale by a factor $\lambda$ the linear dimension of a square or a cube, for example, the corresponding area and volume, scale as $\lambda^2$ and $\lambda^3$, because $D=2$ and $D=3$ are their respective dimensions. On the other hand, if we rescale the linear dimension of a fractal object by the same factor, its spatial content scales by the factor $\lambda^\alpha$, where $D=\alpha$ is its anomalous dimension.  As we show in the present paper, we may expect that, if a physical phenomenon or process takes place on an underlying self-similar structure, then all the scales have to manifest in its evolution, along with the relative scaling factor. In other words, the functional spatial and/or temporal dependence, ruling the phenomenon at hand, should manifest scaling properties related to the ones characterizing the underlying geometry.\\
The connection between fractals and the emergence of non integer order power laws, appears immediately clear, once the self-similar structure of the power type functions is pointed out. If we zoom out by a factor $\lambda$ the time scale in the power law $f(t)=t^\alpha$, the rescaled function reads as: $f(\lambda t)=\lambda^\alpha\,t^\alpha$, that still is the same power law, rescaled by the factor $\lambda^\alpha$. The self-similarity and scaling property of the $t^\alpha$, are then exactly the same as for a fractal object of anomalous dimension $D=\alpha$. From what said, we should expect that the order of the observed power laws, ruling the evolution of the natural phenomena, are strictly related to the dimension of the underlying geometry (as well as to the physics of the problem at hand).

Because of such considerations, aim of the paper is, in first instance, to convince the reader (with the aid of relevant examples) that the previous arguments hold true for whichever phenomenon or process taking place on an underlying fractal geometry, because of the generality of the geometrical and physical considerations on which they are based. Moreover, the main goal of the paper is to perform a step forward and fill the gap between fractals and fractional calculus. Basing on the relationship established between power-laws and fractal geometries, we will show that, once we assume sufficiently small the amplitude of the loads acting on the system, then a fractional differential equation naturally comes up, ruling the dynamics of the system itself, whose order turns out to be related to the dimension of the underlying geometry.

In the following sections, we will show the validity of these considerations studying the time evolution of the flow of a viscous fluid through a porous medium. As expected, a power-law time dependence for the overall flux has been found, whose order is related to the anomalous dimension of the pattern along which we suppose the fluid flows, and to a coefficient related to the physical model by which the motion of the fluid itself is described.

\section{The main model}
\begin{figure}[!h]\centering
\includegraphics[width=2.5cm]{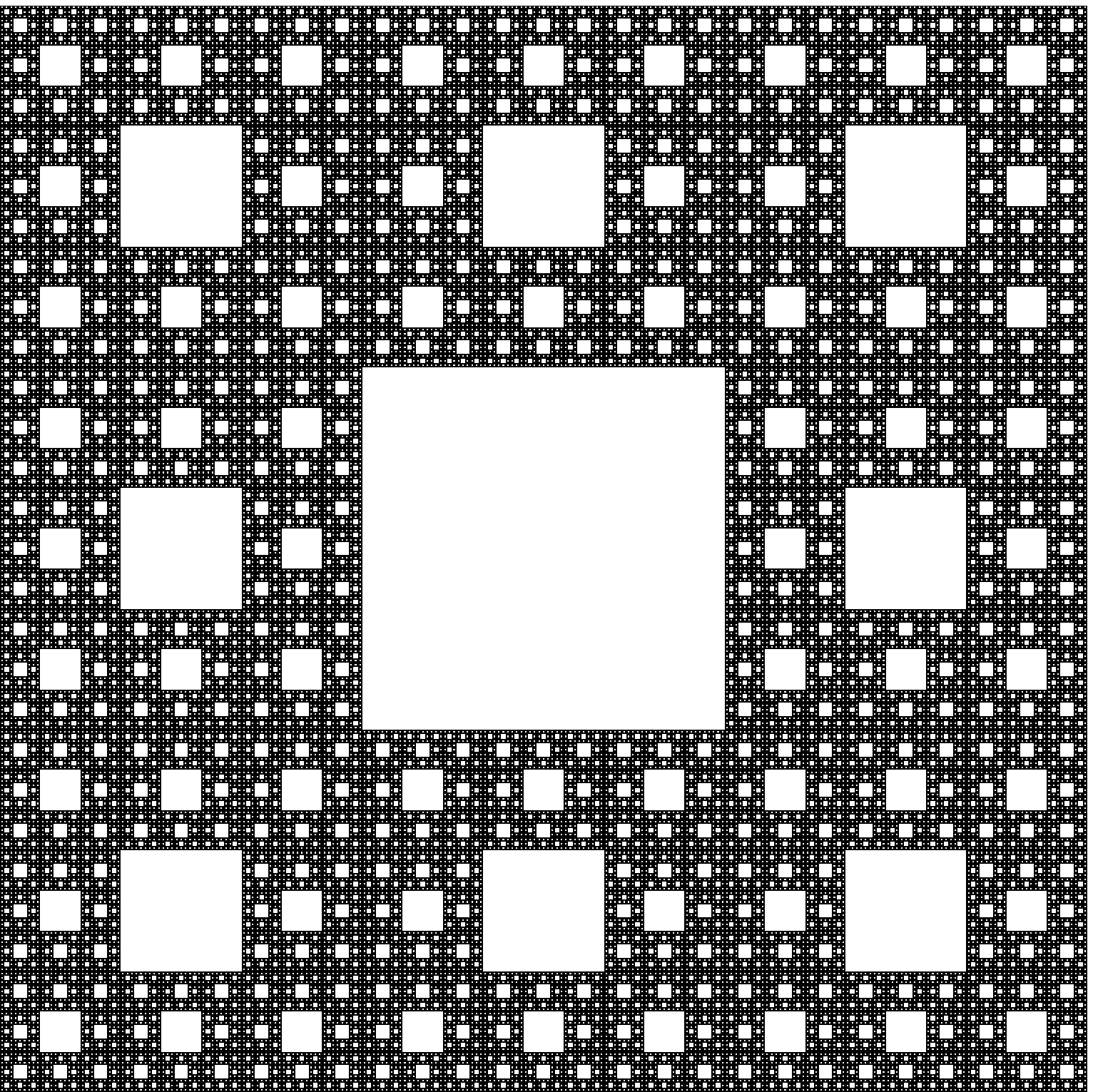}
\qquad
\includegraphics[width=3cm]{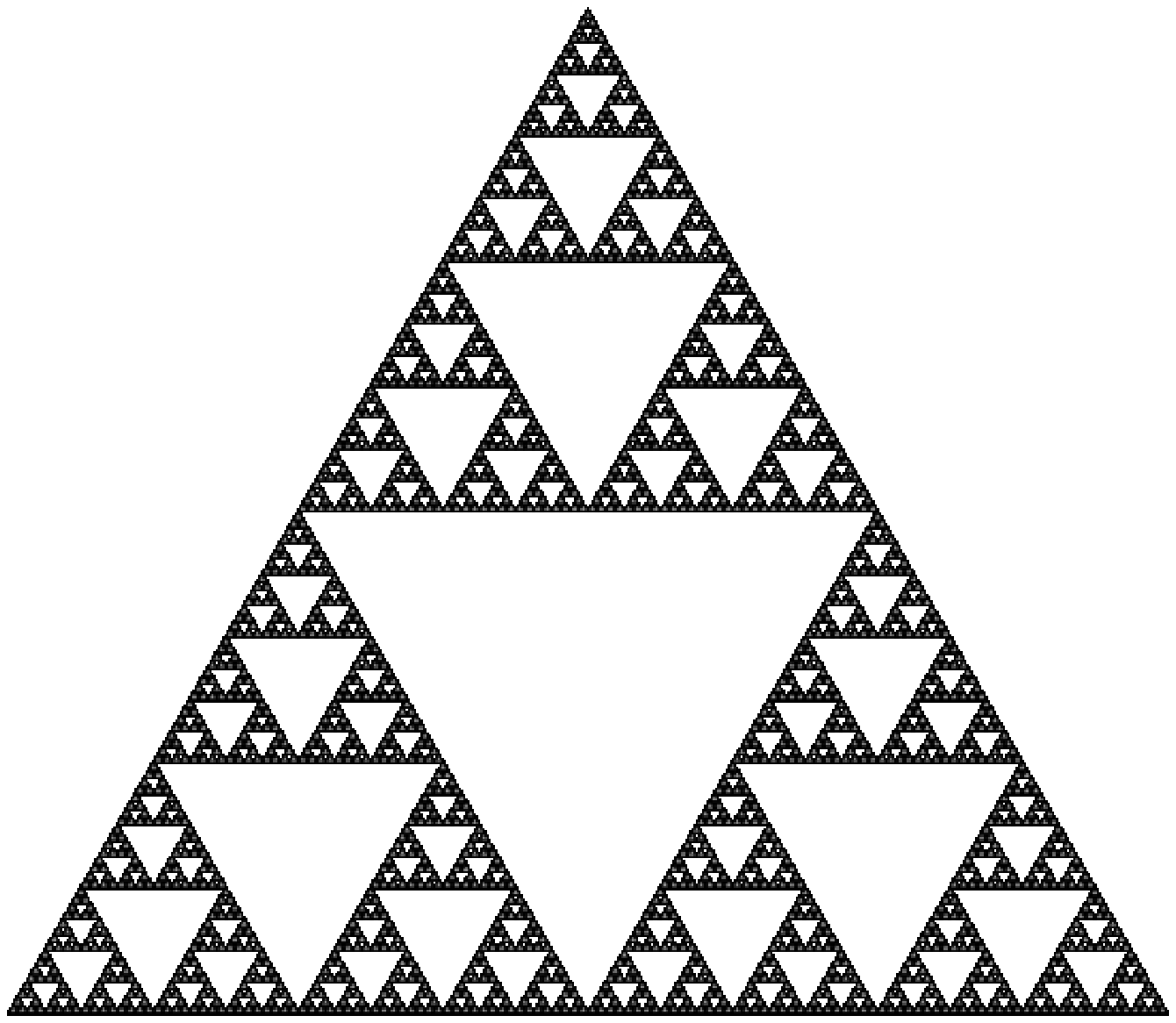}
\caption{The square and triangular Sierpinski carpets}
\label{fig:2}
\end{figure}

In this and in the following sections, we prove the emergence of a power law time dependence in the flux of the viscous fluid flowing through the porous medium. We find an order for the power-law, related to the anomalous dimension of the fractal shape by which we model the pattern of the streams along which the fluid flows. Although we don't give any rigorous mathematical demonstration of the validity of this statement independently on the physical situation at hand, we are confident that, because of the generality of the geometrical and physical considerations on which the following arguments are based, the results derived in the paper hold true for whichever process evolving on an underlying fractal structure.
\begin{figure}[!h]\centering
\includegraphics[width=7cm]{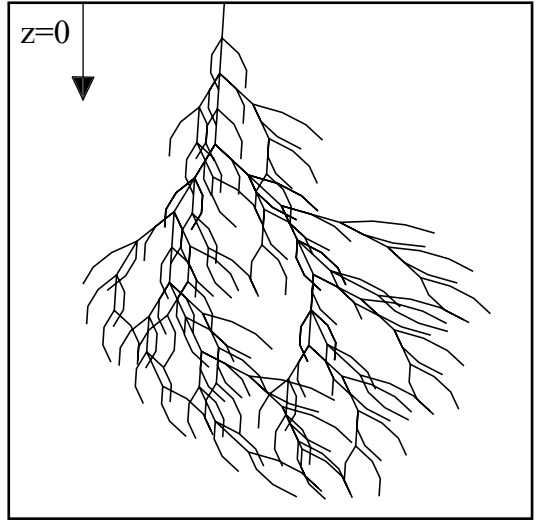}
\caption{Sketch of a Lichtenberg figure, used to model the pattern according to which the fluid flows through the porous medium}
\label{fig:1}
\end{figure}
\vspace{0mm}
Fractal-like structures are widely observed in nature, and the shapes of lightening, trees, river's estuary, ferns, etc. are examples. Since that,  the ``tree'' pattern shown in fig.\ref{fig:1}, known in literature as Lichtenberg figure \cite{Takahashi19791, lichtenberg1778}, from the name of the physicist who originally discovered it in the discharging of insulating material, is likely to describe the flow of a viscous fluid through a porous medium. Such pattern consists of a main stream, branching off in different secondary streams, with progressively smaller diameter, as we go downwards from $z=0$. In the limit $z\to\infty$, i.e. carrying forward the branching process up to infinity, the cross section of such pattern reveals a fractal stochastic shape, in which each stream is surrounded by secondary streams. For simplicity's sake, in our calculations, we will replace such complex structure with an hollowed brick, in which all the streams are equally long, and distributed according to a deterministic fractal shapes such as the square or triangular Sierpinsky carpets shown in fig.\ref{fig:2}. This simplifying hypothesis doesn't affect the physical meaning and the generality of our arguments, because we have simply replaced the complex structure in fig.\ref{fig:1}, by a much simpler one, but still having the key scaling properties characteristic of a fractal shape. 
In what follows we start by setting up the general calculations leading to an expression for the time evolution of the out coming flow, based only on geometrical considerations, and so valid for whichever physical situation we are interested in. In the next sections, we will address the following specific problems: i) in section 3, we start by proving the natural appearance of a power-law in ruling the time evolution of the flux through the brick defined above, that we take as a model for the pattern according to which the streams, along which the fluid flows, are distributed. To this aim, we study the phenomenon of percolation of the fluid, assuming it is subjected only to the gravitational and friction forces. ii) we show in section 4 that, if a time-dependent load acts on the system, that we might identify in a net pressure history applied to the brick, then, once the hypothesis of linear regime is assumed, a constitutive law of non integer order emerges in relating such pressure to the time evolution of the flow through the brick.

Let us start by defining the notation that will be used through the paper. We indicate the number of $n-$order streams and the total area of the corresponding cross-section by $N_n=N^n$ and $S_n=\chi \ell_n^2$ respectively, being $\chi$ a form factor characteristic of the cross section, $N$ the branching factor, and $\ell_n=\ell_0/\epsilon^n$ its characteristic length (in which $\ell_0$ is the quantity relative to the primary stream, and $1/\epsilon$ is the scaling factor of the fractal shape considered). We indicate in general with $v_n(t)$ the velocity of the fluid, flowing along the $n-$th order stream, which depends on the physical model used to describe the motion of the fluid as well as on the scaling law of the fractal at hand. We will carry out the calculations in the continuum limit for which $n\to\alpha\in\mathbb{R}$, that means to treat the discrete fractal as a continuum one, so that the above defined quantities become functions of the variable $\alpha$. Pushing the branching process up to infinity, the out coming flow $\phi(t)$ of the fluid, flowing through the discrete fractal shape, has the form:
\vspace{0mm}
\begin{equation}
    \phi(t)=\sum_{n=0}^\infty {S_n N_n v_n(t)}
\label{eq:1}
\end{equation}
\vspace{0mm}
in which the sum runs over all the orders of the fractal shape, according to which the streams are distributed. The same equation, in the continuum limit, reads as:
\vspace{0mm}
\begin{equation}
    \phi(t)=\int_{0}^{\infty}{d\alpha\,S(\alpha) N(\alpha) v(\alpha,t)}
\label{eq:2}
\end{equation}
\vspace{0mm}
Such equation can be rewritten in therms of only the (inverse of the) scaling factor $\epsilon$, and the anomalous dimension $D=\frac{\ln{N}}{\ln{\epsilon}}$ of the fractal at hand, as:
\vspace{0mm}
\begin{equation}
    \phi(t)=\chi\ell_0^2\int_{0}^{\infty}{d\alpha\,\epsilon^{(D-2)\alpha}v(\alpha,t)}=\chi\ell_0^2\,v_{\text{eff}}(t)
\label{eq:3}
\end{equation}
\vspace{0mm}
having defined the effective velocity
\vspace{0mm}
\begin{equation}
    v_{\text{eff}}(t)=\int_{0}^{\infty}{d\alpha\,\epsilon^{(D-2)\alpha}v(\alpha,t)}
\label{eq:4}
\end{equation}
\vspace{0mm}
Since the velocity of the flow depends on the order $\alpha$ only through the corresponding characteristic length $\ell(\alpha)=\ell_0\epsilon^{-\alpha}$, it is useful to perform the change of variable
\vspace{0mm}
\begin{equation}
    x=\epsilon^{-\alpha};\quad dx=-x\ln\epsilon\,d\alpha;\quad \alpha=0\to x=1;\quad \alpha=+\infty\to x=0
\label{eq:5}
\end{equation}
\vspace{0mm}
and so to write eq.\eqref{eq:4} as
\vspace{0mm}
\begin{equation}
    v_{\text{eff}}(t)=\frac{1}{\log\epsilon}\int_{0}^{1}{dx\,x^{(1-D)}v'(x,t)}
\label{eq:6}
\end{equation}
\vspace{0mm}
being $v'(x,t)$ the functional form of the velocity, in the new variable $x$. The effective velocity is so given by the sum of all the stream's velocities, weighted by the power-type kernel $x^{1-D}$. It is worth to stress that such kernel depends only on the dimension $D$ of the shape by which we describe the cross section of the fractal brick and, as it is evident from eq.\eqref{eq:4}, in the integer limit $D\to 2$ the effective velocity is given simply by the sum of the velocities of the different streams , because in this limit the cross section of the brick scales as a bi-dimensional object, and so the total area relative to the streams of whichever order, is equal to the area of the main one.

Up to now, nothing has been said about the physics involved, that is encoded in $v(x,t)$, but we obtained eqs.\eqref{eq:4} and \eqref{eq:6} only basing on geometrical considerations. In the next sections we will introduce physics in our model, describing the flow along each stream by using the Bernoulli equation, and modelling the losses by the Darcy-Weisbach law. As anticipated, we will show that in both the situations considered, a power law for the overall flux naturally comes up, whose order is related to the anomalous dimension of the pattern according to which the streams are distributed (i.e. to the fractal dimension of the cross section of the brick).

\section{Power laws from fractals: the percolation problem}
Let us start by considering the case in which the brick is saturated with the fluid, at rest for $t\leq 0$, and let us study the time evolution of the out coming flux for $t>0$. The only forces that we suppose acting on the system in such configuration, are the gravitational and the friction ones. Assuming the area of each stream small enough so that the friction forces are predominant respect to the gravitational ones, the velocity of the fluid can be considered, with good approximation, constant along each stream, so that we can write:
\vspace{0mm}
\begin{equation}
\    v'(x,t)=V(x,h)\L[H(t)-H(t-T(x,h)\R]
\label{eq:7}
\end{equation}
\vspace{0mm}
where $H(x)$ is the unit step function, and $V(x,h)$ and $T(x,h)$ are respectively the constant value of the velocity, and the discharging time of the $x$-th stream, that of course depend on the thickness $h$ of the brick. Because of physical considerations, we should expect that smaller is the diameter of the stream, i.e. less is the parameter $x$ (or larger is $\alpha$), smaller is the value $V(x,h)$ of the velocity of the flow because more relevant is the effect of the friction forces and, consequently, larger is the discharging time $T(x,h)$.  It is worth to stress that the assumption of small cross section, that allowed us to write the velocity along the streams as in eq.\eqref{eq:7}, doesn't affect our results form a qualitatively point of view, but has only been made in order to solve the integral in eq.\eqref{eq:6} in a closed form. Substituting eq.\eqref{eq:7} into eq.\eqref{eq:6}, we get the following expression for the effective velocity:
\vspace{0mm}
\begin{equation}
    v_{\text{eff}}(t)=\frac{1}{\log\epsilon}\int_{0}^{1}{dx\,x^{(1-D)}V(x,h)\L[H(t)-H(t-T(x,h)\R]}
\label{eq:8}
\end{equation}
\vspace{0mm}
In the limit of small cross sections, we can Taylor expand the $x-$dependence of $V(x,h)$ and so write, to the lowest significant order, $V(x,h)=\bar{V}(h) x^{\beta}$ and $T(x,h)= h/V(x,h)=\L(h/\bar{V}(h)\R) x^{-\beta}=\bar{T}(h)x^{-\beta}$ (with $\beta$ a coefficient defined from a proper physical model or experimentally deduced, and $\bar{T}(h)=h/\bar{V}(h)$). Substituting such relations in eq.\eqref{eq:8}, we then have
\vspace{0mm}
\begin{equation}
\begin{split}
   v_{\text{eff}}(t)&=\frac{\bar{V}(h)}{\log\epsilon}\int_{0}^{1}{dx\,x^{(1-D+\beta)}\L[H(t)-H(t-T(x,h)\R]}
\end{split}
\label{eq:9}
\end{equation}
\vspace{0mm}
Such integral can be carried out noting that the integrand is not null only in the time interval $0<t<T(x,h)$, obtaining:
\begin{equation}
v_{\text{eff}}(t)=\frac{\bar{V}'(h)}{\beta\gamma\log\epsilon}\,t^{-\gamma} \qquad\text{for}\quad t>0
\label{eq:10}
\end{equation}
in which we indicated $\gamma=1+(2-D)/\beta$ and $\bar{V}'(h)=\bar{V}(h)^{{1-\gamma}}\,h^{\gamma}$. Eq. \eqref{eq:10} proves that the time evolution of the overall flux is ruled by a power-law, whose order is related to the anomalous dimension $D$ of the cross section of the brick, and to a coefficient $\beta$, characteristic of the model used to describe the motion of the fluid along the streams.

\begin{figure}[!h]\centering
\includegraphics[width=6cm]{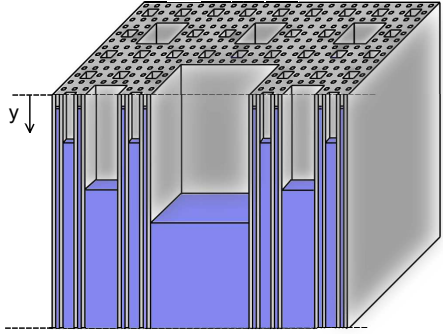}
\caption{Discharging of the viscous fluid through the hollowed brick, defined in the text as a simplifying model for the Lichtenberg figure reported in fig.\ref{fig:1}.}
\label{fig:3}
\end{figure}

Let us introduce now a physical model into the general arguments outlined so far. As anticipated, we model the energy losses in the flow due to the friction forces, by using the Darcy-Weisbach equation, according to which such losses, after a length $L$, per unit mass of the fluid , are equal to:
\vspace{0mm}
\begin{equation}
    \Delta E_{\text{loss}}=K \frac{v^2}{2g}\frac{L}{d}
\label{eq:11}
\end{equation}
\vspace{0mm}
where $d$ is the ``hydraulic diameter'' of the stream's section, and $K$ is a dimensionless coefficient called the Darcy friction factor. The value of this factor can be deduced from the Moody's diagram and, in the assumption of laminar flow (that is the limit we are working on), it is equal to $K=64/Re$, being $Re=\frac{\rho v d}{\nu}$ the Reynolds number of the flow.
The Bernoulli equation, by which we describe the motion of the fluid, particularized for the problem at hand (see fig.\ref{fig:3}), reads as
\vspace{0mm}
\begin{equation}
    \frac{v^2}{2}+\L(64\frac{\nu}{\rho\, v\, d}\R)\frac{v^2}{2g} \frac{\L(h-y\R)}{d}=g\L(h-y\R)
\label{eq:12}
\end{equation}
\vspace{0mm}
having indicated by $y$ the distance of the surface of the column of fluid, in the generic stream, from the top of the brick, by $\rho$ and $\nu$ respectively the density and the dynamic viscosity of the fluid, and by $g=9.81 \text{m/s}^2$ the standard gravitational acceleration. Solving this equation for the velocity $v$ of the fluid, we obtain:
\vspace{0mm}
\begin{equation}
    v(d,y)=-32\frac{\nu}{\rho\,g\,d^2}(h-y)+\sqrt{1024 \L(\frac{\nu}{\rho\,g\,d^2}\R)^2(h-y)^2+2g(h-y)}
\label{eq:13}
\end{equation}
\vspace{0mm}
so that the time instant at which the surface of the column of water is at the distance $y$ from the top, is formally given by:
\begin{equation}
    t(y)=\int_0^y{\frac{dy'}{v(y')}}
\label{eq:14}
\end{equation}
\vspace{0mm}

\begin{figure}[!h]\centering
\includegraphics[width=8cm]{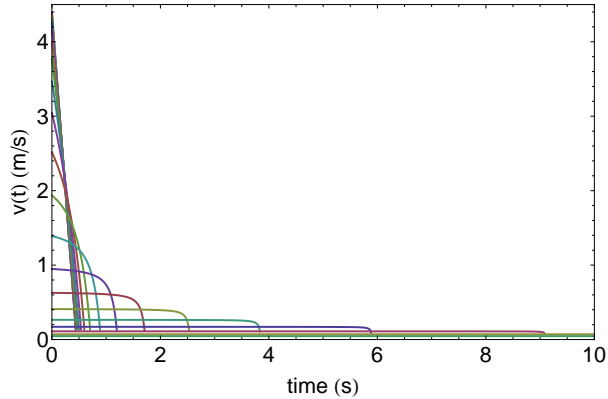}
\caption{Velocity as a function of time, for pipes from the order $\alpha=0$ to $\alpha=4$, with a step $\Delta\alpha=0.2$ }
\label{fig:4}
\end{figure}

\begin{figure}[htbp]
\centering%
\subfigure[Square Sierpiknsi carpet]%
{\includegraphics[width=8cm]{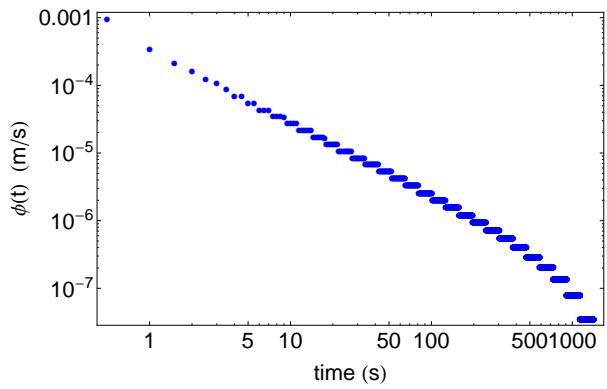}\label{fig:5a}}\qquad\qquad
\subfigure[Triangular Sierpiknsi carpet]%
{\includegraphics[width=8cm]{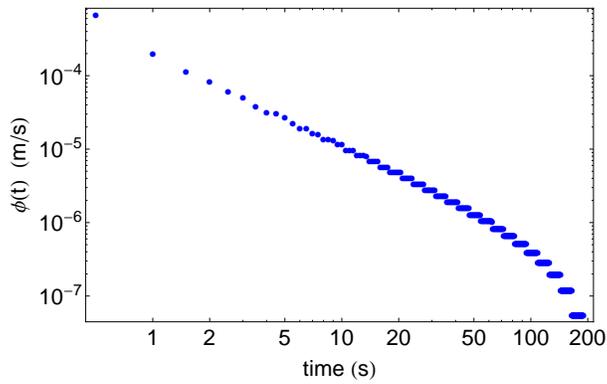}\label{fig:5b}}
\caption{Overall flux as a function of time\label{fig:5}}
\end{figure}

Because of the assumption of small area of the streams, and so small hydraulic diameters, the eq.\eqref{eq:13} can be approximated to the lowest order as:
\vspace{0mm}
\begin{equation}
    v(y)\simeq \frac{g^2 \rho d^2}{32\nu}=\xi\frac{g^2 \rho \ell^2}{32\nu}
\label{eq:15}
\end{equation}
\vspace{0mm}
being $\xi$ a coefficient relating the hydraulic diameter and the characteristic length $\ell$ of the streams. Remembering that such length scales as $\ell(x)=\ell_0 x$, we can identify in eq.\eqref{eq:15} the parameters $\bar{V}(h)=\xi\frac{g^2\rho\ell_0}{32\nu}$ and $\beta=2$, previously defined.
Inserting in eq.\eqref{eq:10}, the Hausdorff fractal dimension $D=1.8928$ for the square Sierpinski carpet, and $D=1.5849$ for the triangular Sierpinski carpet, together with the value $\beta=2$ calculated above, we obtain that, for such geometries of the cross section of the brick, the fluxes are proportional respectively to: $\phi_{\text{Sq}}(t)\sim t^{-1.0536}$ and $\phi_{\text{Tr}}(t)\sim t^{-1.2075}$, being the former relative to the square carpet and the latter to the triangular carpet.\\
In order to verify these results, and so the validity of eq.\eqref{eq:7}, we have calculated numerically and directly from the eqs.\eqref{eq:4}, \eqref{eq:13} and \eqref{eq:14}, the flux as a function of the time, by using the values $\epsilon=3$, $N=8$ that define the square carpet and $\epsilon=2$, $N=3$ that define the triangular one. We assumed the side $\ell_0=1\, \text{cm}$ for the main stream, and a thickness $h=1\,\text{m}$ of the brick. Furthermore we considered the density $\rho=1000\,\text{Kg/m}^3$, and the viscosity $\nu=0.001\,\text{Pa}\cdot \text{s}$ for the fluid. Solving numerically the system composed by eqs.\eqref{eq:13} and \eqref{eq:14}, we computed the velocity of the flow as a function of time, for fractal orders ranging between $0<\alpha<4$, with a step $\Delta\alpha=0.2$. The result is shown in fig.\ref{fig:4}, and reveals the goodness of the approximation we made, in writing $v(x,t)$ in terms of step functions.
Considering now the wider range from $\alpha=0$ to $\alpha=6$, and a step $\Delta\alpha=0.1$, we obtained the time evolution of the overall flux from the brick, shown in log-log scale in figs.\ref{fig:5a} and \ref{fig:5b} respectively for the square and triangular carpet. Apart for a deviation in the long time domain, due to the finite number of fractal orders used in the calculation, the time dependence is clearly a power-law. By fitting these numerical results, disregarding the data relative to the long time domain, we get that the flux, for the square and the triangular carpet respectively, is proportional to $\phi_{\text{Sq}}(t)\sim t^{-1.057}$ and $\phi_{\text{Tr}}(t)\sim t^{-1.24}$, results that are very close to the ones calculated from eq.\eqref{eq:11}.
\newpage

\section{The rise of fractional operators in the linear regime: the seeping problem}
In the previous section we dealt with the problem of percolation of a fluid through a porous medium, that has been modelled by the fractal hollowed brick. We showed that, as the flow takes place on every scale of the fractal, then the overall out coming flux evolves in time according to a power-law, whose order keeps memory of the scaling properties of the underlying fractal geometry, as well as of the physical model by which the motion of the fluid has been described. Basing on this remarkable result, our goal now is to give evidence of the occurrence of fractional operators in the equations ruling the evolution of such flow, once a pressure history is applied on the medium. We will show in particular  that it happens once the response of the system to the forcing action is linearised around a fixed value, condition that allow us to take advantage of the Boltzmann superposition principle. This is exactly the same hypotesis on which the well known theory of linear viscoelasticity is based. As long as the value of the load $F$ applied to a specimen of viscoelastic material is sufficiently small in fact, then the non linear dependence of the elongation $u$ on the load itself, found by Nutting in 1921 in his famous experiment  \cite{Nutting1921}, and equal in general to
\vspace{0mm}
\begin{equation}
    u=a\,t^n\,F^m
\label{eq:16}
\end{equation}
\vspace{0mm}
can be linearised around the value $F=0$. In eq.\eqref{eq:16} $m$ and $n$ are parameters characteristic of the material, that vary with the temperature, while the coefficient $a$ depends on the type of experiment performed. Again, in this assumption the Boltzmann superposition principle is now valid so that, as we consider the specimen subjected to a load history, the time evolution of the elongation is ruled by a linear fractional differential equation. By following the same arguments here described for the viscoelasticity, we will linearise the dependence of the flux on the pressure applied to the brick, obtaining a relation between these two quantities, of non integer order.

\begin{figure}[htbp]
\centering%
\subfigure[Square Sierpiknsi carpet]%
{\includegraphics[width=8cm]{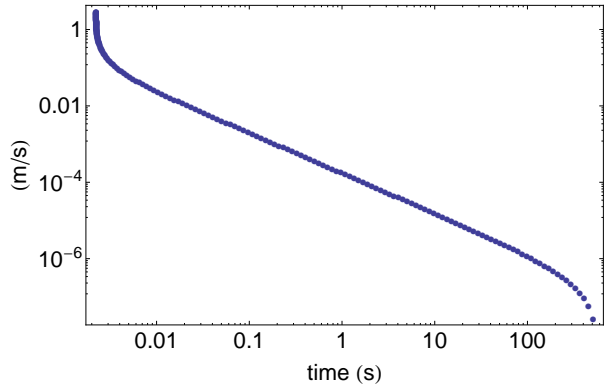}\label{fig:6a}}\qquad\qquad
\subfigure[Triangular Sierpiknsi carpet]%
{\includegraphics[width=8cm]{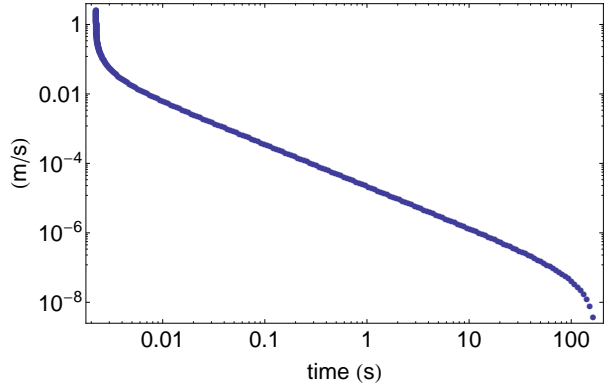}\label{fig:6b}}
\caption{Overall flux as a function of time\label{fig:6}}
\end{figure}

Since it is not significant in this context, we disregard in what follows the effect of the gravity and of any other volume forces acting on the fluid, assuming it subjected only to the friction forces and to the net pressure acting on the two sides of the brick (that we assume empty for $t\leq 0$). At the time instant $t=0$, we imagine to connect it to a reservoir, in which the fluid is at rest and whose pressure is assumed to be constant in time. Our first aim is then to derive an expression for the time evolution of the flux seeping through the brick, as a function of the net pressure $p$ applied to the brick.\\
Again, in the assumption of sufficiently small area of the streams, the friction forces are predominant respect to the pressure ones and so the velocity in every streams can be assumed constant. In this hypothesis we can then write:
\vspace{0mm}
\begin{equation}
    v'(x,t)=V(x,p)H(t-T(x,p))
\label{eq:17}
\end{equation}
\vspace{0mm}
where $V(x,p)$ and $T(x,p)$ are, respectively, the velocity of the fluid at the lower side of the brick, and the time necessary for for the fluid itself to cover the distance $h$, equal to the thickness of the brick. Inserting eq.\eqref{eq:17} into the eq.\eqref{eq:6} for the effective velocity $v_{\text{eff}}(t)$, we get
\vspace{0mm}
\begin{equation}
\begin{split}
    v_{\text{eff}}(t)&=\frac{1}{\log\epsilon}\int_{0}^{1}{dx\,x^{(1-D)}V(x,p)H(t-T(x,p))}
\end{split}
\label{eq:18}
\end{equation}
\vspace{0mm}
Describing again the motion of the fluid by using the Bernoulli equation, and the energy losses by the Darcy-Weisbach law, because of the hypothesis of small cross section of the streams, we can Taylor expand the expressions for the velocity and the discharging time, obtaining, to the lowest significant order, $V(x,p)=\bar{V}(p) x^{\beta}$ and $T(x,p)= h/V(x,p)=h/\bar{V}(p) x^{-\beta}=\bar{T}(p)x^{-\beta}$, with $\bar{V}(p)=\bar{V}_0 \,p=\L(\frac{\rho g \ell_0^2}{32\nu h}\R)\, p$, $\bar{T}(p)=\bar{T}_0/p=\L(\frac{32\nu h^2}{\rho g \ell_0^2}\R)\frac{1}{p}$ and again with $\beta=2$. Since the unit step function in eq.\eqref{eq:18} is  not null for $t>T(x,p)=\bar{T}(p)x^{-\beta}$, condition that is satisfied for $x>\L(\frac{t}{\bar{T}(p)}\R)^{-1/\beta}$, the integral can be solved in a closed form, giving 
\vspace{0mm}
\begin{equation}
    v_{\text{eff}}(t)=\frac{\bar{V}(p)}{\log\epsilon\,\beta\gamma}\L[1-\L(\frac{t}{\bar{T}(p)}\R)^{-\gamma}\R]
     \qquad \text{for} \quad t>\bar{T}(p)
\label{eq:19}
\end{equation}
\vspace{0mm}
where, as before, $\gamma=1+(2-D)/\beta$. We obtained again a power-law ruling the evolution of the flux, whose order is exactly the same as the one obtained in the previous section. This remarkable result confirms that the order of the power law depends only on the anomalous dimension of the underlying fractal geometry and on the model used to describe the physics, and is independent on the particular situation at hand. As done in the previous section, we verified the validity of eq.\eqref{eq:17}, by calculating numerically and directly form the Bernoulli equation, the flux as a function of time. We considered the range of fractal orders between $\alpha=0$ and $\alpha=10$, with a step $\Delta\alpha=0.05$. We isolated the time dependent part of such solution, that is reported in fig.\ref{fig:6a} and \ref{fig:6b} in log-log scale. Disregarding the short and long time domains, in which a deviation from a power-law is evident and due to the finite number of the fractal orders used in the calculation, we fitted the data, obtaining the time dependences for the flux $\phi_{\text{Sq}}\sim t^{-1.0536}$ and $\phi_{\text{Tr}}\sim t^{-1.2075}$, respectively for the square and triangular Sierpinksi carpets, that again are very close to the one obtained from eq.\eqref{eq:19}.

So far we have calculated the response of the system (in terms of the overall flux of the viscous fluid) to a step-like excitation. In section 2, such excitation was given by the gravitational forces that, switched on at the time instant $t=0$, enforced the fluid to flow through the medium. In this section instead, we dealt with the case in which the flow is driven by a constant net pressure applied to the medium, again for $t>0$. We found in both cases, a power-law for such response, whose order is related to the dimension of the fractal shape by which we modelled the pattern of the streams along which the fluid flows through the medium. Basing on these results, the last fundamental step we want to perform, and that represent the principal aim of the paper, is to give evidence of the emergence of fractional differential equations, in ruling the response of the system, once it is subjected to a load history. Because of the linearity of fractional operators, we prove the validity of this statement, once the non linear dependence of the response on the amplitude of the forcing action, is properly linearized. In particular, in the present context, we prove the emergence of fractional operators in the constitutive law relating the overall flux of the fluid and the net pressure history applied to the brick, once the eq.\eqref{eq:19}, that represent the response of the system to a step-like forcing action, is linearized around a fixed value $p=p_0$. To this aim, let us assume such pressure history varying in a small region of pressure values around $p_0$. This assumption allows us to linearize eq.\eqref{eq:19} in the neighbourhood of $p=p_0$. Disregarding the constant pressure contribute, and focusing on the effect due to $\Delta p(t)=p(t)-p_0$, by exploiting the Boltzmann superposition principle, we can write the effective velocity in the form:
\vspace{0mm}
\begin{equation}
\begin{split}
v_{\text{eff}}(t)&=C_1\L[\Delta p(t)+C_2\int_0^{t-\bar{T}(p_0)}{\frac{\Delta\dot{p}(\tau)}{(t-\tau)^{\gamma}}d\tau}\R]  \qquad \text{for} \quad t>\bar{T}(p_0)
\end{split}
\label{eq:25}
\end{equation}
\vspace{0mm}
where we defined the coefficients
\vspace{0mm}
\begin{equation}
\begin{split}
	C_1&=\frac{\bar{V}_0}{\beta \gamma \log\epsilon}\\
	C_2&=\L(\gamma-1\R)\bar{T}_0^{\gamma}p_0^{-\gamma}
	\end{split}
\label{eq:26}
\end{equation}
\vspace{0mm}
In the eq.\eqref{eq:25} for the effective velocity, an integral term appears, that looks like a fractional operator. It is easy to show that integrating it by parts, and assuming the initial conditions $\Delta p(t=0)=0$ and $\Delta \dot{p}(t=0)=0$, the constitutive law relating the effective velocity (and so the overall flux) and the pressure history applied to the brick, is a differential equation of not integer order that reads, again for $t>\bar{T}(p)$, as:
\vspace{0mm}
\begin{multline}
v_{\text{eff}}(t)=C_1\Delta p \left(t-\bar{T}(p_0)\right)+\frac{C_2}{\gamma-1}\Delta \dot{p} \left(t-\bar{T}(p_0)\right)\\
-\Gamma\left(2-\gamma\right)\frac{C_2}{\gamma-1}\L(\prescript{C}{}{\mathbf{D}}_{\bar{T}(p_0)}^{\gamma} \Delta p\R)\L(t-\bar{T}(p_0)\R)
\label{eq:27}
\end{multline}
\vspace{0mm}
where $\Dc{p}{\gamma}$ is the Caputo's definition of fractional derivative of order $\gamma$. The eq.\eqref{eq:27} is the fundamental result we were looking for. It proves that, in the hypothesis of linear regime, a fractional differential equation naturally comes up in ruling the evolution of a natural phenomenon or process, taking place on an underlying fractal geometry, whose order is related to the anomalous dimension of the latter. We derived such result by studying the physics of the flow of a viscous fluid through a fractal shape. Anyway the considerations made, that lead us to such remarkable result, are general and so hold true whichever is the physical problem at hand.

This result gives us a deep insight into the world we live in. We know that differential operators of integer orders, as well as integer order ordinary or partial differential equations, are peculiar of  smooth geometries, on which all the physical theories developed up to now, from the newton laws of mechanics, to the Einstein theory of gravitation, or quantum mechanics, are based. On the other hand, according to the results obtained in this paper, the success of fractional calculus in modelling physical phenomena, reveals an underlying fractal nature of reality. Because of that, we are confident that in the next future, the fractal geometry will have a leading rule in gaining a better and more complete understanding of physics and of the others natural sciences, hopefully clarifying some of the issues by which are still affected.

\section{Conclusions}
In this paper, we showed that a relation between fractional calculus and fractal geometry exists, which is intimately related to the physical origins of the power-law long memory and hereditary properties observed in many natural phenomena, and that are the characteristic feature of fractional operators. With the aid of the relevant example of a viscous fluid seeping through a porous medium, that we modelled as a fractal shaped hollowed brick, we showed that a power-law naturally comes out, ruling the evolution of a physical phenomenon, every time it takes place on an underlying fractal geometry. Trace of the scaling properties of the underlying fractal geometry, exhibits in the order of such power-law that, accordingly, we found to be strictly related to its anomalous dimension, and to a coefficient, characteristic of the model by which we describe the physics of the phenomenon at hand.\\
Linearizing the non linear dependence of the response of the systems at hand to a proper forcing action we proved, by exploiting  the Boltzmann superposition principle, that a fractional differential equation naturally comes up in ruling its evolution. The order of such equation turned out to be related to the anomalous dimension of the geometry on which the phenomenon takes place.

\section*{Acknowledgement}
S.B. acknowledges support from EPSRC CM-DTC.

\newpage
\bibliography{FractionalOperatorsBiblio}
\bibliographystyle{unsrtnat}
\include{FractionalOperatorsBiblio}

\end{document}